# Single-molecule junctions map the interplay between electrons and chirality

Anil-Kumar Singh[1], Kévin Martin[2], Maurizio Mastropasqua Talamo[2], Axel Houssin[2], Nicolas Vanthuyne[3], Narcis Avarvari[2*], and Oren Tal[1*]

[1]*Department of Chemical and Biological Physics, Weizmann Institute of Science, Rehovot 7610001, Israel*
[2]*Univ Angers, CNRS, MOLTECH-Anjou, SFR MATRIX, F-49000 Angers, France*
[3]*Aix Marseille Univ, CNRS, Centrale Marseille, UAR 1739, FSCM, Chiropole, Marseille, France*
[*] Corresponding authors

**Abstract**

The interplay of electrons with a chiral medium has a diverse impact across science and technology, influencing drug separation, chemical reactions, and electronic transport[1-30]. In particular, such electron-chirality interactions can significantly affect charge and spin transport in chiral conductors, ranging from bulk semiconductors down to individual molecules. Consequentially, these interactions are appealing for spintronic manipulations. However, an atomistic mapping of the different electron-chirality interactions and their potential for spintronics has yet to be reached. Here, we find that single-molecule junctions based on helicene molecules behave as a combined magnetic-diode and spin-valve device. This dual-functionality is used to identify the coexistence of different electron-chirality interactions at the atomic-scale. Specifically, we find that the magnetic-diode behavior arises from an interaction between the angular momentum of electrons in a chiral medium and magnetic fields, whereas the spin-valve functionality stems from an interaction between the electron's spin and a chiral medium. The coexistence of these two interactions in the same atomic-scale system is then used to identify the distinct properties of each interaction. This work uncovers the different electron-chirality interactions available at the atomic level. The found concurrent existence of such interactions can broaden the available methods for spintronics by combining their peculiar functionalities.

## Main

The interactions between electronic angular momentum, whether in a spin or orbital form, and a chiral medium hold diverse fundamental and practical implications. For example, these interactions are directly associated with molecular recognition, charge transfer in biosystems, chemical reactions, drug purification, and, foremost, with electronic transport in chiral conductors across all relevant scales and dimensions, down to individual molecules[1-30]. As a fundamental symmetry-related subject with broad impact, the details of these interactions have been subjected to extensive research[1-44]. However, an atomistic picture of the interplay between electronic angular momentum and a chiral medium remains elusive, along with its full potential for spintronic manipulations.

In the last two decades, a large set of phenomena related to electron transport and transfer in chiral conductors has been studied experimentally. The observed phenomena have been typically attributed to one of two general effects: the chiral-induced spin selectivity (CISS)[1,5-9,12,14,15,19-23,25-43] and the electrical magnetochiral anisotropy (EMCA, sometimes denoted as eMChA)[2-4,10,11,13,16-18,21,23,24]. In the former case (CISS; Fig. 1a), an electron moving in a chiral conductor experiences an effective magnetic field with an orientation that depends on the conductor's chirality and the direction of the electron's velocity. The effective magnetic field interacts with the electronic magnetic moment, promoting the transport of electrons with one spin direction (either parallel or antiparallel to the velocity) and suppressing the transport of electrons with the opposite spin direction. Thus, for a given chirality and current direction, the electronic current is dominated by one spin population. In the latter case (EMCA; Fig. 1b) that to date has not been identified in atomic-scale systems, the angular momentum of an electron moving in a chiral system is affected by the chiral landscape. The interaction between the resulted angular momentum and an external magnetic field, parallel or antiparallel to the electron's velocity, promotes or suppresses electron transport. In this case, the conductor's resistance is decreased or increased by the EMCA effect depending on the chirality of the system, the current direction, and the external magnetic field orientation.

In this work, we reveal the simultaneous occurrence of the EMCA and CISS effects at the atomic-scale and characterize their properties at the limit of quantum transport. Specifically, we find that single-molecule junctions based on helicene molecules behave as a merged magnetic-diode and spin-valve device, due to a coexistence of the EMCA and CISS effects. The distinct nature of these effects is unveiled by their different response to applied magnetic fields, and electrodes composed of metals with different spin-orbit coupling (SOC). We find no apparent coupling between the EMCA and CISS effects and identify the conditions in which their magnitude is equal. We uncover an unknown response of the EMCA effect to SOC, and the absence of a similar response for the CISS effect. This important observation can limit the range of relevant theoretical models for the two effects. Our work maps the different contributions that dominate the interplay between electrons and a chiral medium at the atomic scale. The found coexistence of the CISS and EMCA effects at this scale presents opportunities for a broader range of spintronic manipulations in miniaturized systems, leveraging the different nature of each effect.

We use single-molecule junctions prepared in a break-junction setup at 4.2 Kelvin (Fig. 1c[44,45]). The junctions include a Ni electrode as a source or drain for spin-polarized current, a counter electrode made of Au, Ag, or Cu, and an unprecedented 2,2'-dithiol-[6]helicene (helicene hereafter) as a chiral molecular bridge (see Supplementary Section 1 for synthesis and characterization). The choice of the molecule was motivated by the well-known affinity of thiol groups for the coinage metals and by the robust helical chirality of the helicene framework[46]. Before the molecules are introduced, the contact between the electrode tips is repeatedly broken and reformed in sub-atomic precision. This process wets the Ni tip with the softer metal of the counter electrode to have two atomic-scale apexes made of the softer metal[44]. Next, the helicene molecules are introduced into the cold junction by in-situ sublimation from a local source during repeated junction breaking and squeezing[45]. We use either the *P*-enantiomer of helicene with a clockwise helicity or the *M*-enantiomer with an anticlockwise helicity (Fig. 1, insets). The described junction fabrication and the following measurements are done in a cryogenic temperature and ultra-high vacuum conditions that minimize unwanted contaminations. See details in Methods and Supplementary Section 2.

**Results**

**Current-voltage curves under magnetic fields, asymmetry and magnetoconductance**

Figures 1d-i present histograms and average current in absolute values, as a function of applied voltage (|I|-V curves) measured for hundreds of molecular junction realizations. Before each measurement, the two electrode apexes are squeezed against each other and then stretched to reform a new molecular junction in order to sample the span of different molecular junction configurations. Separate sets of |I| -V measurements were performed for molecular junctions based on *M* (Figs. 1,d,e) and *P* (Figs. 1,g,h) enantiomers. During the measurements, a constant magnetic field of +2 or −2 Tesla (T) was applied to align the Ni magnetization parallel or antiparallel to the junction's axis. Consequentially, a spin-polarized current is generated at a finite voltage with a dominant population of spins aligned either antiparallel or

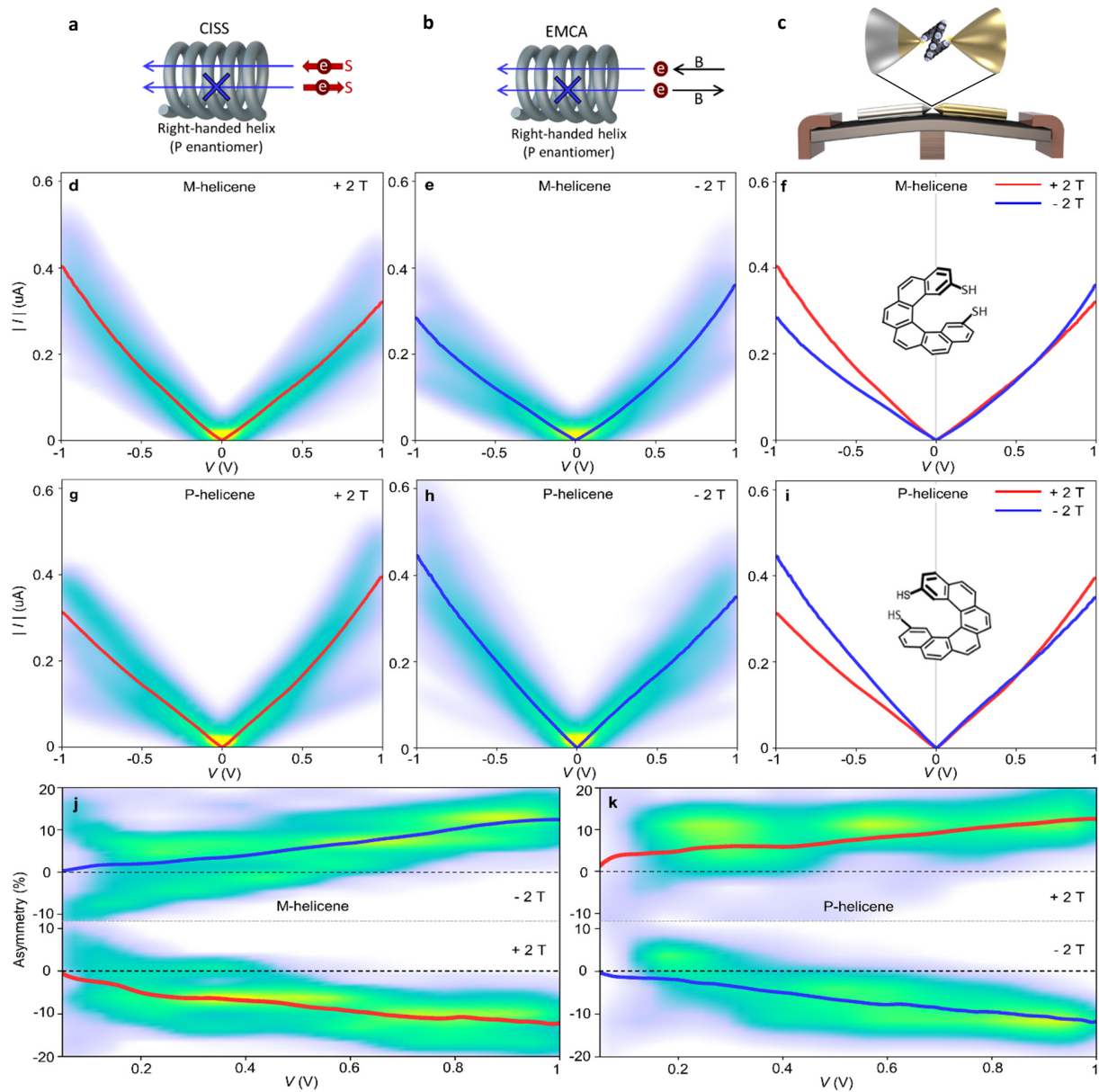

**Fig. 1: Current-voltage analysis of helicene molecular junctions under magnetic fields. a,** Illustration of chiral-induced spin selectivity (CISS). **b,** Illustration of electrical magnetochiral anisotropy (EMCA). Here, the helix indicates a chiral conductor, red circles - electrons (e), red arrows - spin (S), blue arrows - electron transport directions, black arrows - magnetic field (B) directions. **c,** Illustration of a break-junction setup and a helicene molecular junction. **d,** Histogram and an average of current in absolute values as a function of voltage (|I|-V curves) for Ni(Au)/*M*-helicene/Au junctions under +2 T magnetic field, parallel to the junction. **e,** The same under -2 T magnetic field antiparallel to the junction. **f,** Average of absolute value of current as a function of voltage for Ni(Au)/*M*-helicene/Au junctions under parallel and antiparallel +2 T and -2 T magnetic fields. **g-i,** The same as (d-f) but for Ni(Au)/*P*-helicene/Au junctions. The standard error of the current [$(standard\ deviation)/\sqrt{\#\ of\ curves}$] in D to I is smaller than the curve width. **j,k,** Asymmetry as a function of voltage magnitude for Ni(Au)/*M*-helicene/Au (j) junctions under the mentioned opposite magnetic fields. Asymmetry is defined as $Asymmetry = 100 \cdot [|I(+V)| - |I(-V)|]/[|I(+V)| + |I(-V)|]$. **k,** The same but for Ni(Au)/*P*-helicene/Au junctions. The number of examined molecular junctions (also the number of I-V curves) in each case varies between 251 to 377.

parallel to the junction's axis. To have a better comparison between these cases, the average |I| -V curves for opposite magnetic fields, are presented together in Figs. 1,f,i, for each enantiomer (refer to Fig. S6 for a polar I-V presentation and Fig. S7 for individual I-V curves). Interestingly, the |I|-V curves are asymmetric, revealing current rectification or diode-like behavior. Namely, the current magnitude is different for a positive and negative voltage. For a given enantiomer, the asymmetry is inverted when the magnetic field direction is reversed (blue versus red in Figs. 1f,i). Moreover, for a given magnetic field (e.g., blue curves in Figs. 1,f,i), the asymmetry is inverted when opposite molecular chirality (*P* or *M*) is used. This is quantitatively summarized in Figs. 1,j,k, by asymmetry histograms and average asymmetry as a function of voltage (see Fig. 1 caption for "asymmetry" definition). The observed inversion of asymmetry when opposite chirality or magnetic field direction are used rules out the possibility of an asymmetric junction structure as the source of asymmetry in the |I|-V curve. Thus, the origin of the diode-like behavior of the helicene junctions is clearly related to the application of magnetic fields and the molecule's chirality.

**CISS and EMCA in current-voltage curves and magnetoconductance**

The identified current rectification stands in contrast to the characteristics of I-V measurements reported in previous experiments related to the CISS effect across a wide range of systems. These systems include a chiral conducting medium positioned between ferromagnetic and non-ferromagnetic electrodes[6,19-21,26,27,29]. Irrespective of the diverse architectures and materials used, in all these cases the reported response of the I-V curves to opposite magnetization or chirality is symmetrical in the following sense. The current magnitude in one curve is always larger than that of the other curve, regardless of voltage polarity (Fig. 2a and Fig. S8a). This behavior was ascribed to the injection of spins with opposite orientations at positive and negative applied voltages[47]. We include in this definition also previously reported asymmetric I-V curves resulting from uneven voltage drops across an asymmetric junction structure, where the current magnitude may differ for opposite voltages. However, it consistently remains larger for a specific chirality and magnetic field direction when compared to their opposite counterpart[6,29,48]. In contrast to the findings related to the CISS effect, the EMCA effect induces a suppression of resistance for one voltage polarity and an enhancement in resistance for the opposite voltage polarity[2-4]. Specifically, for a given chirality and magnetic field orientation, the contribution to resistance (or conductance, which is 1/resistance) by the EMCA effect changes its sign, depending on the current direction. This is translated into current rectification and an asymmetric I-V curve[11,24,49-51], as depicted in Fig. 2b and Fig. S8b, showing partial resemblance to our measurements (e.g., Fig. 2,e,f).

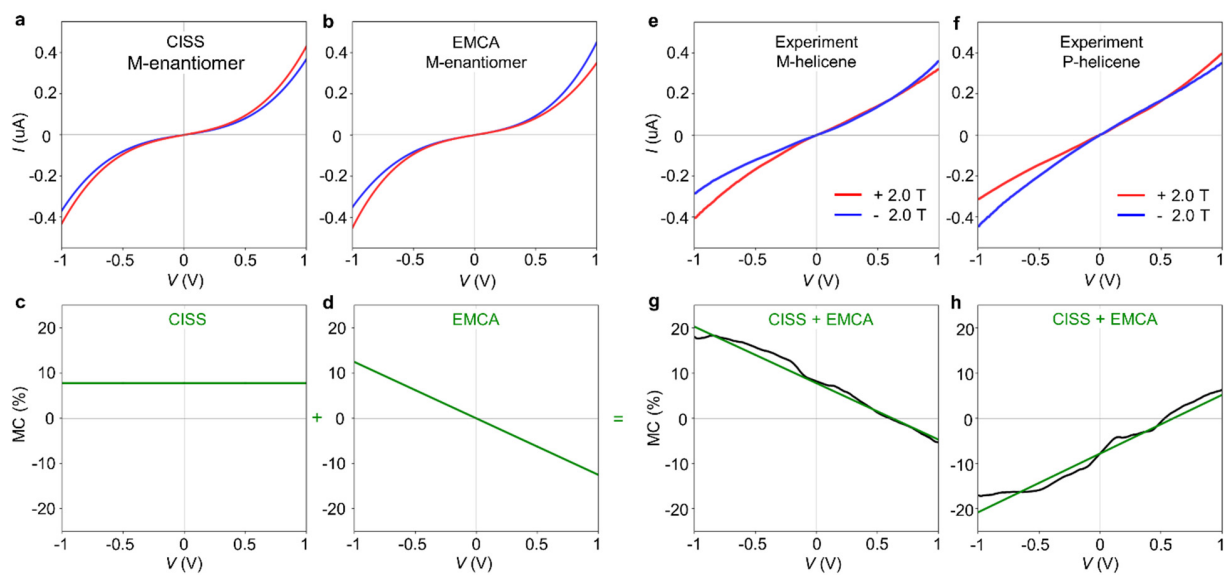

**Fig. 2: Magnetoconductance in view of the CISS and EMCA effects. a,b,** Simulated I-V curve for the CISS effect (a), and the EMCA effect (b). **c,d,** Simulated MC for the CISS effect (c), and the EMCA effect (d). **e,f,** Measured average I-V curve for hundreds of Ni(Au)/*M*-helicene/Au junctions (e) and Ni(Au)/*P*-helicene/Au junctions (f) at +2 T and -2 T applied magnetic fields. The standard error of the current in (e-f) is smaller than the curve width. **g,** Measured average MC (black) based on data from (e) for *M*-helicene junctions. The green curve represents a fit to the black measured curve, serving as the basis for generating the simulated I-V and MC curves in panels (a-d). **h,** Measured average MC based on data from (f) for *P*-helicene junctions. The green curve is not a fit to the data in (h), but a mirror inversion of the fit for the measured MC of the *M*-helicene junctions seen in (g). Note the agreement between the inverted curve based on data obtained in a set of experiments for the *M*-helicene junctions and the data obtained in independent set of experiments for the *P*-helicene junctions. The number of examined molecular junctions in each case varies between 251 to 377.

The expected manifestations of the two effects can be clearly seen in magnetoconductance (MC), defined as: $\mathrm{MC} = [G\uparrow(V) - G\downarrow(V)]/[G\uparrow(V) + G\downarrow(V)]$, where $G = I/V$ is the conductance. For the CISS effect, a symmetric MC is expected, as shown in Fig. 2c, since the conductance (curve's slope) at both voltage polarities is larger for one I-V curve compared to the other (Fig. 2a). In this case, the sign of MC depends on the chirality of the system, resulting a positive MC for the *M*-enantiomer (Fig. 2c) and a negative MC for the *P*-enantiomer (Fig. S8c). For the EMCA effect, as seen in Fig. 2b, at a positive voltage the conductance is larger for one I-V curve compared to the other, while the situation is reversed for a negative voltage. This is translated to an antisymmetric MC that changes signs at zero voltage (Fig. 2d). Furthermore, the sign of the MC slope depends on the chirality of the system: a negative slope for the *M*-enantiomer (Fig. 2d) and a positive slope for the *P*-enantiomer (Fig. S8d). The MC obtained from the measured I-V curves (Fig. 2,g,h, black) reveals deviations from the expected behavior of either the CISS or EMCA effects. The experimental MC curves are tilted with negative and positive slopes for the *M*- and *P*- enantiomers, respectively, consistent with the behavior ascribed to the EMCA effect. However, the transition from positive to negative MC does not occur at zero voltage, contrary to the expected characteristic of the EMCA effect. Ignoring the fine MC structure that may originate from the specific electronic structure of the junction, the green graph in Fig. 2g captures the essence of this behavior for the *M*-enantiomer. It represents a linear combination of the expected MC curves for the CISS and EMCA effects seen in Fig. 2,a,b. Namely, the general behavior of the probed MC can be explained by the contribution of both effects: the CISS accounts for the *MC shift*, while the EMCA introduces the *MC tilting*, which stems from the discussed asymmetry. The green graph in Fig. 2g is a fit to the measured data for *M*-helicene junctions. Interestingly, the green graph in Fig. 2h is the same graph from Fig. 2g yet with an inverted slope sign, revealing a remarkable agreement with the measured data for *P*-helicene junctions in evidently independent experiments (Supplementary Section 3).

**The influence of magnetic field magnitude**

As mentioned, the EMCA effect arises in the presence of an external magnetic field. In contrast, the CISS effect is not expected to be influenced by such fields, except for a negligible influence from Zeeman splitting. Figures 3,a-i present the $|I|$-V curves, asymmetry, and MC in three different magnetic fields. Here, magnetic field above 2 T were considered to ensure magnetization saturation even at the Ni atomic apex, and higher fields than 4 T were avoided due to expected contributions from high-order corrections to the EMCA effect[3,7,24]. While there are evident differences between the curves, to get quantitative information we first focus on the asymmetry at 1 V as a function of magnetic field as presented in Fig. 3j, which reveals a clear dependance. The asymmetry is proportional to the conductance difference between positive and negative voltage (Supplementary Section 4), which is expected to be linear and reduced to zero in the absence of magnetic field for the EMCA effect[2,3]. Applying a linear fit (red) to the six data points reveals through extrapolation that the asymmetry vanishes in the absence of a magnetic field as expected for the EMCA effect. In practice, this behavior cannot be directly observed since the Ni electrode induces a finite magnetic field even when the external field is nullified.

Plotting in Fig. 3k the detected MC shift (MC at zero voltage) as a function of magnetic field magnitude reveals that it is not sensitive to the field, as expected for the CISS effect (see Supplementary Section 5 and Figs. S9a,b for an alternative MC shift analysis). The asymmetry response to magnetic fields and the lack of detected influence of the field on the MC shift, support the earlier conclusion that the asymmetry is an outcome of the EMCA effect, while the observed MC shift is a consequence of the CISS effect. The zero MC at a positive voltage signifies a specific point where the influence of the EMCA and the CISS effects is equal and opposite, resulting in MC nullification. This point shifts to a lower voltage with an increase in magnetic field, as expected in view of the EMCA response to magnetic field strength. Interestingly, the linear response of the asymmetry to magnetic fields and the absence of any detected effect of magnetic fields on the MC shift suggest that within our experimental sensitivity there is no coupling between these manifestations of the CISS and EMCA effects.

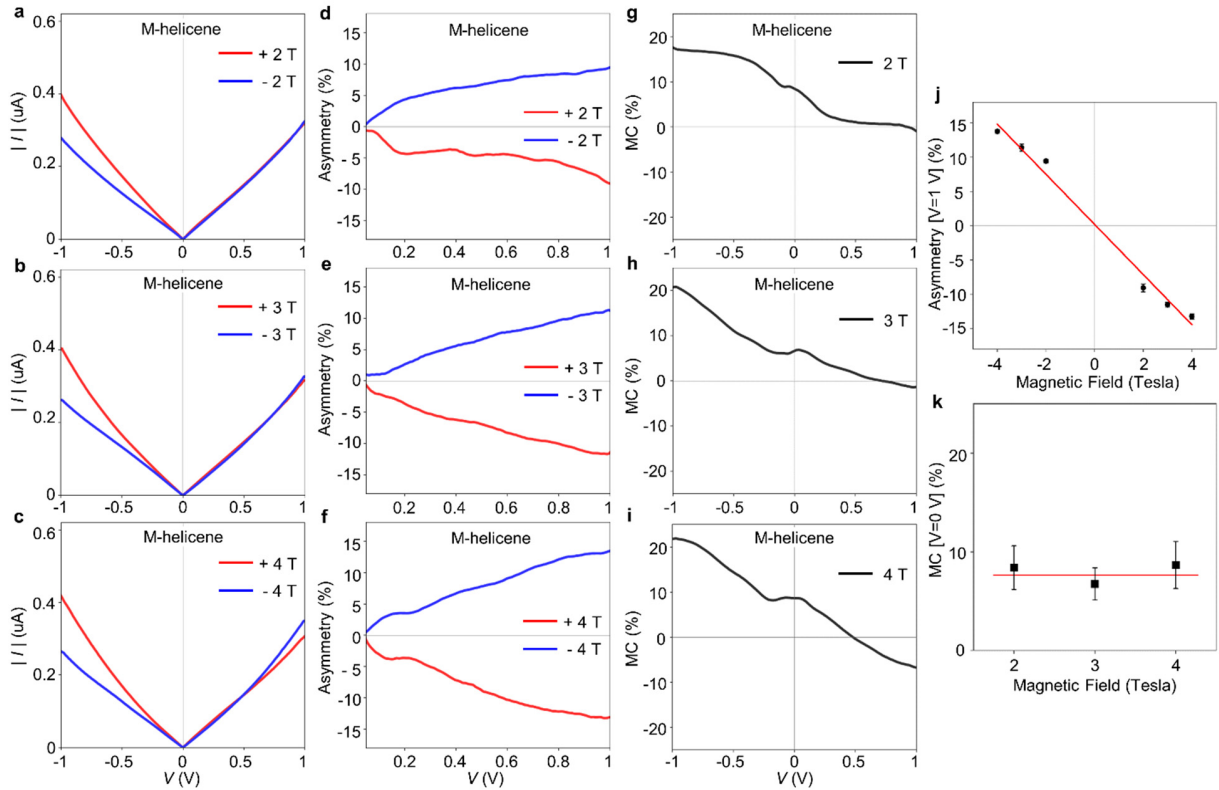

**Fig. 3: Asymmetry and MC analysis at different magnetic fields. a-c,** Average current (in absolute values) as a function of applied voltage for Ni(Ag)/*M*-helicene/Ag junctions at different applied magnetic fields. The standard error of the current is smaller than the curve width. **d-f,** Average asymmetry as a function of applied voltage magnitude at different applied magnetic fields. **g-i,** Average MC as a function of applied voltage at different magnitudes of magnetic field. **j,** Asymmetry at 1 V as a function of magnetic field (k) MC shift (MC at zero voltage) as a function of magnetic field magnitudes. The error bars for asymmetry and MC indicate the experimental uncertainty in view of the standard deviation of the measured currents. The number of examined molecular junctions in each case varies between 372 to 634. We study the response to magnetic field magnitudes using junctions based on Ag rather than Au. This choice is motivated by the tendency of Au to form atomic chains, which enhances result variability and complicate the analysis, especially when minor trends should be carefully detected. See Fig. S10 for corresponding |I|-V and asymmetry histograms.

**The influence of metal electrodes with different spin-orbit coupling**

Examining the response of the two effects to a common variable can further test their coexistence, while providing insights into the distinct nature of each effect. Below, we consider the influence of different non-ferromagnetic electrodes made of Cu, Ag, and Au, having in mind their different SOC with increasing magnitude: Cu<Ag<Au[52]. In all three cases, the molecular junctions are characterized by a similar conductance around $5·10^{-3}$ $G_o$ (Fig. S5). Figures 4a-c, present the measured average |I|-V curves for the three cases when applying parallel and antiparallel magnetic fields for *M*-helicene junctions (see Fig. S12 for a similar analysis of *P*-helicene junctions). The |I|-V response to magnetic fields varies among junctions based on the three different metals. Examining in Figs. 4d-f the resulted asymmetry, we find a monotonous increase in its magnitude. Fig. 4J summarizes the total asymmetry (sum of positive and negative asymmetry magnitudes) at 1 V for the three different metals, where a larger asymmetry is observed for metals that exhibit a larger SOC. Focusing in Fig. 4,g-i on MC, the increased tilt observed along the Cu, Ag, and Au series is another manifestation of the mentioned asymmetry trend. However, the MC shift presented in Fig. 4k as MC at zero voltage is not sensitive to the metal type (see Supplementary Section 5 and Figs. S9c,d for an alternative MC shift analysis).

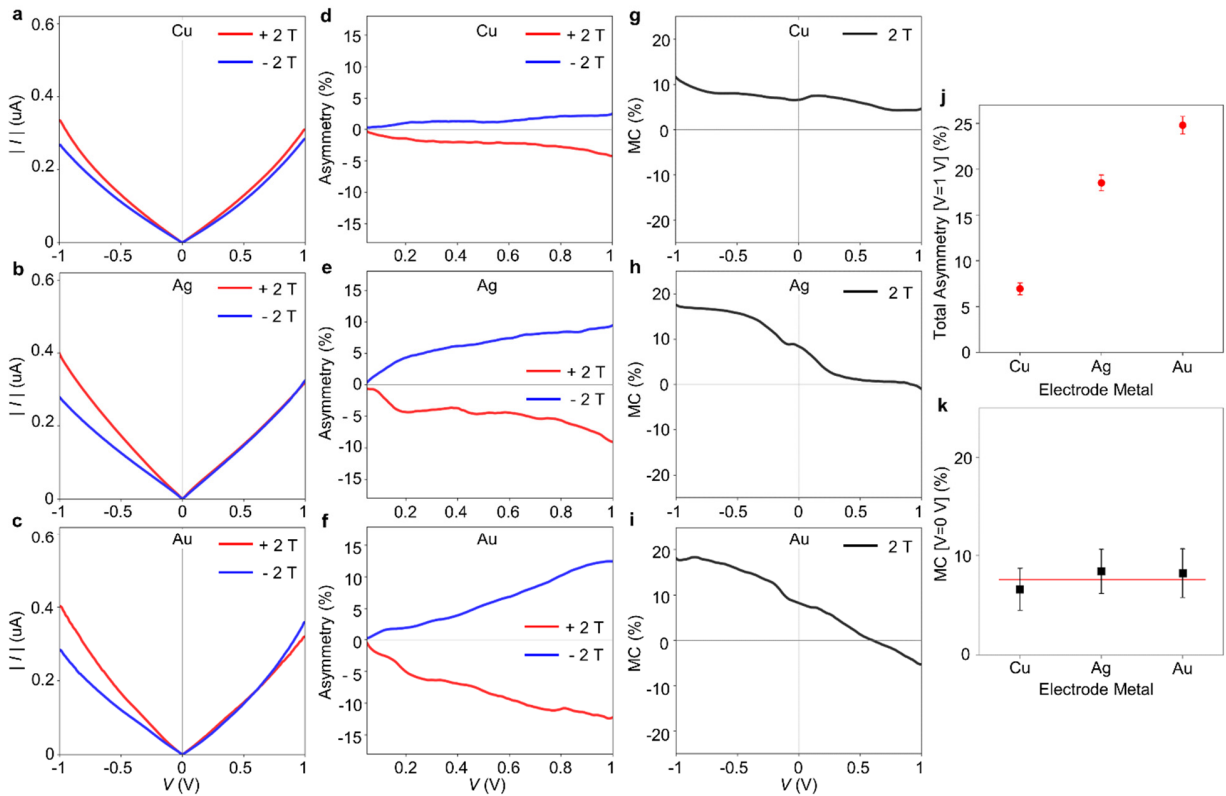

**Fig. 4: Asymmetry and MC response to different metal electrodes. a-c,** Average current (in absolute values) as a function of applied voltage for Ni(X)/*M*-helicene/X junctions, where X is Cu (a), Ag (b) and Au (c). The standard error of the current is smaller than the curve width. **d-f,** Average asymmetry as a function of applied voltage magnitude for the same junctions as in (a-c), respectively. **g-i,** Average MC as a function of applied voltage for the same junctions as in (a-c), respectively. **j,** Total asymmetry at 1 V for junctions based on different metals. **k,** MC shift (MC at zero voltage; black squares) for junctions based on different metals. The red curve represents the average value. The number of examined molecular junctions in each case varies between 316 and 443. The error bars for asymmetry and MC indicate the experimental uncertainty. See Figs. S10,a-c, Fig. S11, and Fig 1d,e for |I|-V and asymmetry histograms for the three cases. See Fig. S12 and Fig. S13 for a similar analysis of *P*-helicene based junctions.

The different response of asymmetry and MC shift to the metal type strengthen the conclusion that they stem from two different effects, in accordance with the accumulated indications presented above for the coexistence of the CISS and EMCA effects. The observed increase in asymmetry along the set of Cu, Ag, and Au provides a first systematic indication for a possible influence of SOC on the EMCA effect. This provides guidelines for a theoretical examination of the role of SOC in the EMCA effect, a dimension that is currently absent. The association of the CISS effect with MC shifts and the absence of a clear MC shift response in Fig. 4k, suggest that at the limit of the measurement uncertainty, the CISS effect is not sensitive to the electrode's SOC or other variance between the used Au, Ag and Cu electrodes, in agreement with[14,16], where similar metals were used. Note that the slightly lower response for the Cu based electrode is observed both here and in[14,16]. This contrasts the observations reported in ref. 27, where the use of an Al substrate led to a significantly lower MC compared to an Au substrate. We can point to one difference in the mentioned comparative analyses: all the mentioned metals have distinct SOCs, but Cu, Ag, and Au have dominant s frontier orbitals at the Fermi energy, in contrast to Al with dominant p on top of s frontier orbitals. These may indicate on the sensitive role of the substrate's atomistic properties in determining the spin-dependent transport via metal-chiral molecule interfaces. Generally, if the CISS effect is indeed independent of the metal's SOC as observed here, it narrows down the range of theoretical explanations pertinent to the CISS effect in similar systems.

**Discussion and conclusions**

The clear indications for the EMCA effect in the examined single-molecule junctions raise a question regarding the conditions in which this effect can be observed at the atomic or molecular scale. Previous I-V measurements in chiral molecular junctions were typically performed as a function of magnetization orientation of one of the electrodes in order to explore the CISS effect. In these studies, a planar multi-molecular geometry or a scanning probe microscope configuration were usually adopted, where a flat ferromagnetic thin film was used as a central component of one of the electrodes (e.g., refs. 6,20,29). These structures have essentially a negligible intrinsic magnetic field, ignoring the film's edges. In another example, the molecule was placed away of the ferromagnet[26]. Beyond a sizable magnetic field, a significant current density can also enhance the EMCA response[50]. While high current densities are not typical for multi-molecular junctions, they are expected for single-molecule junctions[29]. In fact, the combination of both: a sizable magnetic field, and a high current concentration, are met in our single-molecule junction experiments. In our setup, one of the electrodes is made of bulk Ni with an intrinsic magnetic field, and the junction is subjected to external fields of at least 2 T. Furthermore, the current concentration is around several $10^7 A/cm^2$. Therefore, we expect that the EMCA effect will be seen in chiral single-molecule junctions with similar current concentrations and magnetic fields. The finite CISS response found in our study near zero voltage, may seemingly violate the constraints set by time reversal symmetry. According to the latter, the CISS should be nullified within the linear response regime[34,35,37,40,43]. Yet, our observations well agree with former measurements of a finite CISS response at low applied voltages across ferromagnet-based two terminal devices[53], thus providing guiding lines for theoretical descriptions of the CISS effect.

To conclude, in this work chiral single-molecule junctions are used to map the interplay of electrons and chirality at the atomic scale. This electron-chirality interaction dominates charge and spin transport in chiral materials. We uncover the simultaneous occurrence of the CISS and EMCA effects at the atomic scale, seen as a combined magnetic-diode-spin-valve spintronic functionality. Our analysis reveals no apparent coupling between these effects. Importantly, we find that metallic electrodes with different SOC affect the EMCA response, but not the CISS response. This work provides the first indication for the existence of the EMCA effect at the atomic scale and at the limit of quantum electronic transport. We further reveal an unknown SOC influence on the EMCA effect, offering a starting point for developing an atomistic EMCA theory, which is currently absent. The lack of substrate SOC influence on the CISS effect in electronic transport experiments can be used to narrow down the relevant atomistic mechanisms for this effect. Overall, the coexistence of the CISS and EMCA effects, both of comparable magnitude at the atomic scale, can expand the scope of spintronic functionalities in miniaturized systems by harnessing the unique characteristics of each effect.

**Methods**

**Sample preparation**

The experiments are done in a special version of a mechanical controllable break-junction set-up (Fig. 1c) as described in detail in ref. 43, and briefly here. The samples consist of one electrode made of a Ni wire terminated with a tip and a second counter electrode made of a Au, Ag, or Cu wire also ended with a tip (purity: 99.994%(Ni), 99.998%(Au), 99.997%(Ag), 99.9999%(Cu), diameter: 0.1 mm, length: 6 mm, manufacturer: Alfa Aesar). The two wires are attached to a flexible substrate composed of a phosphor-bronze plate (thickness: 1 mm) covered by an insulating Kapton film (thickness: 100 $\mu m$). Initially, the flexible substrate is bent, and subsequently, the two wires are attached to the bent substrate, with their tips oriented toward each other. Next, the substrate is relaxed to a flat configuration, and the tips are compressed together to form a macroscale contact. This break junction structure is introduced into a vacuum chamber and cooled to 4.2 K. To prepare an atomic-scale junction, the substrate is bent by a piezoelectric element (PI P-882 PICMA) that pushes the substrate at its center against two peripheral stoppers (Fig. 1c). As a result, the tips are pulled apart, and the contact cross-section is gradually reduced until a junction with a single-atom diameter neck is formed between the electrodes. Further extension leads to junction rupture. A fresh atomic junction can be prepared by relaxing the substrate, such that the electrode tips are pressed against each other to establish a multiatomic junction, after which the electrodes are pulled apart again to restore a single-atom junction. This break-make cycle can be iterated for thousands of times such that the Ni electrode is wet by the softer metal of the counter electrode. After characterization of the bimetallic junction (Fig. S5a-c) to verify that the formed junction has now a typical conductance of a bare Cu, Ag, or Au junction as explained in ref. 43, the target molecules are introduced (See Supplementary Section 1 for details about synthesis and characterization of the target molecules). We use a heated local molecular source to sublimate the target molecules into the cold junction, while repeating the break-make cycles. Once the typical conductance of the junction is altered (indicating the presence of molecules in the junction), the sublimation is stopped. Different molecular junctions are prepared by squeezing the electrodes to have a contact with a conductance of ∼30 $G_0$ followed by elongation of the contact up to rupture and the insertion of individual molecules between the electrodes. Repeating this procedure yields ensembles of molecular junctions with a variety of different geometries.

**Conductance-displacement measurements**

Conductance measurements as a function of elongation that provide the conductance histograms seen in Fig. S5 are done in the following way[43]. The junction is elongated at a rate of 20–40 Hz, while the conductance of the junction is measured simultaneously. The junction is biased with a fixed voltage provided by a DAQ card (NI-PCI6221) that is divided by 10 (by a homemade divider) to increase the signal to noise ratio. The resulting current across the junction is amplified by a current preamplifier (Femto amplifier DLPCA 200) and recorded by the DAQ card at a sampling rate of 50–200 kHz. The obtained current values are divided by the applied voltage values to extract the conductance. The interelectrode displacement is found by the exponential dependence of tunneling currents on the separation between the electrodes. The piezoelectric element that is used to bend the sample is driven by the same DAQ card connected to a piezo driver (Piezomechanik SVR 150/1).

**Current-voltage measurements**

Current as a function of voltage measurements are done as follows. Once a molecular junction is formed with a certain conductance in a range between $1 \cdot 10^{-3}$ $G_0$ and $8 \cdot 10^{-3}$ $G_0$ ($G_0 \cong 1/12.9$ (k$\Omega$)$^{-1}$ is the conductance quantum). A variable bias voltage is applied across the junction from the mentioned DAC card and divider. The voltage is swept in a rate of 0.5 V/sec, while the current is measured as mentioned above. During repeated I-V measurements on different molecular junction realizations, a constant magnetic field is applied using a superconducting magnet ($\leq$ 9 T) that provides a magnetic field parallel or antiparallel to the sample wire.

**Data availability**

The data that support the findings of this study are available from the corresponding authors upon request. Source data will be provided with this paper.

**Ethics declarations**

**Competing interests**

The authors declare no competing interests.

**Supplementary Information**

Supplementary Sections 1-6, Supplementary Figures S1-S13, Supplementary Scheme S1, Supplementary Tables S1-S5, and Supplementary References.

Supplementary Information for

# Single-molecule junctions map the interplay between electrons and chirality

Anil-Kumar Singh[1], Kévin Martin[2], Maurizio Mastropasqua Talamo[2], Axel Houssin[2], Nicolas Vanthuyne[3], Narcis Avarvari[2*], and Oren Tal[1*]

[1]*Department of Chemical and Biological Physics, Weizmann Institute of Science, Rehovot 7610001, Israel*
[2]*Univ Angers, CNRS, MOLTECH-Anjou, SFR MATRIX, F-49000 Angers, France*
[3]*Aix Marseille Univ, CNRS, Centrale Marseille, UAR 1739, FSCM, Chiropole, Marseille, France*
[*] Corresponding authors

**Content:**

**Section 1: Synthesis and characterization of the target molecule**

**Section 2: Electron transport measurements of the studied atomic and molecular junctions**

**Section 3: Fit procedure in Fig. 2.**

**Section 4: The relation between asymmetry and conductance difference.**

**Section 5: Magnetoconductance shift extracted by linear fitting**

**Section 6: Additional complementary figures**

**Scheme S1**

**Figures S1 to S13**

**Tables S1 to S5**

**References 1–22**

## Section 1: Synthesis and characterization of the target molecule

Chemicals and instruments

All reagents and chemicals from commercial sources were used without further purification. Solvents were dried and purified using standard techniques. Column chromatography was performed with analytical-grade solvents using Aldrich silica gel (technical grade, pore size 60 Å, 230-400 mesh particle size). Flexible plates ALUGRAM® Xtra SIL G UV254 from MACHEREY-NAGEL were used for TLC. Compounds were detected by UV irradiation (Bioblock Scientific) or staining with iodine, unless otherwise stated.

NMR spectra were recorded with a Bruker AVANCE III 300 ($^1$H, 300 MHz and $^{13}$C, 76 MHz) and Bruker AVANCE DRX 500 ($^1$H, 500 MHz and $^{13}$C, 125 MHz). Chemical shifts are given in ppm relative to tetramethylsilane TMS and coupling constants $J$ in Hz. Residual non-deuterated solvent was used as an internal standard.

Matrix Assisted Laser Desorption/Ionization was performed on MALDI-TOF MS BIFLEX III Bruker Daltonics spectrometer using dithranol, DCTB or α-terthiophene as matrix.

Synthetic procedures

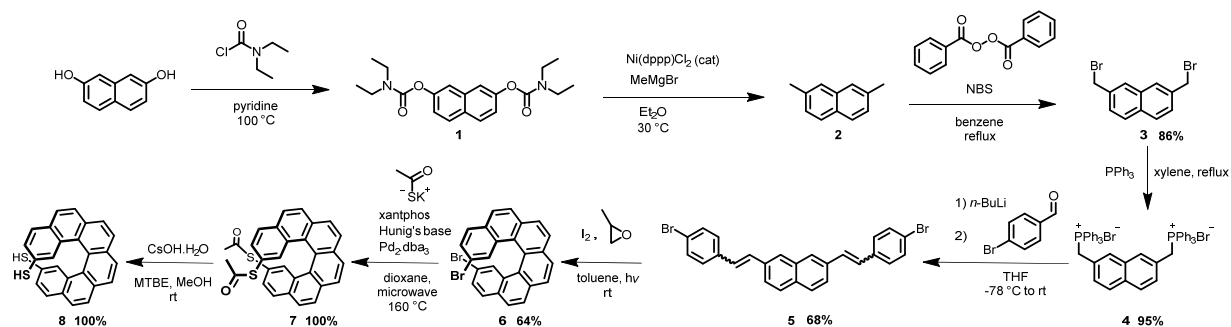

**Scheme S1:** Synthetic pathway for hexahelicene-2,15-dithiol (**8**).

*naphthalene-2,7-diyl bis(diethylcarbamate) (**1**)*

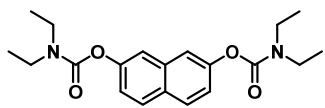

Compound **1** has been synthesized from 2,7-dihydroxy-naphthalene according to the published method (*1*).

*2,7-dimethylnaphthalene (**2**)*

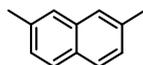

Compound **2** has been synthesized from **1** according to the published method[1].

*2,7-bis(bromomethyl)naphthalene (**3**)*

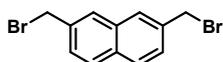 In a Schlenk tube under argon was dissolved **2** (780 mg, 3.22 mmol, 1 eq) in benzene (20 mL), then NBS (578 mg, 25.6 mmol, 1.1 eq) and benzoyl peroxide (105 mg, 0.64 mmol, 0.1 eq) were added in the dark. The mixture was stirred at reflux for 16 h; after reaching rt it was filtered off through a celite® pad. After purification *via* chromatography over silica gel column (PE/DCM, 9/1, Rf = 0.15), compound **3** was obtained as a white solid, 887 mg (86 % yield).

**[1]H NMR** (300 MHz, Chloroform-*d*) δ 7.82 (d, *J* = 8.8 Hz, 4H), 7.52 (d, *J* = 9.8 Hz, 2H), 4.66 (s, 4H).

The spectral data for this compound match those reported in the literature[2].

*(naphthalene-2,7-diylbis(methylene))bis(triphenylphosphonium) bromide (**4**)*

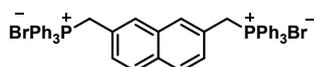 In a 100 mL flask was dissolved PPh$_3$ (5.51 g, 21.02 mmol, 3 eq) in xylene (57 mL), then **3** (2.2 g, 7.01 mmol, 1 eq) was added and the mixture was stirred at reflux for 20 h. After reaching the rt, the precipitate was filtered off and rinsed with cold Et$_2$O to afford (5.57 g (95 % yield) of **4** as a white powder.

**[1]H NMR** (300 MHz, Chloroform-*d*) δ 7.86 – 7.57 (m, 30H), 7.48 (d, *J* = 8.3 Hz, 2H), 7.20 (s, 2H), 7.15 (d, *J* = 8.4 Hz, 2H), 5.53 (d, *J* = 14.3 Hz, 4H).

**[31]P NMR** (122 MHz, Chloroform-*d*) δ 22.90.

The spectral data for this compound match those reported in the literature[3].

*2,7-bis(4-bromostyryl)naphthalene (**5**)*

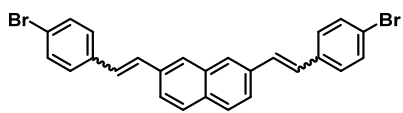 In a 250 mL Schlenk flask under argon was dissolved **4** (5 g, 5.96 mmol, 1 eq) in dry THF (100 mL). At -78 °C, *n*-BuLi (7.83 mL, 12.52 mmol, 1.6 M in hexane, 2.1 eq) was slowly added and the mixture turned from

white to red. After 15 min stirring, the mixture reached rt and stirred for 15 additional min. The mixture was then cooled down at -78 °C and *p*-bromobenzaldehyde (2.2 g, 11.92 mmol, 2 eq) was added. The mixture was stirred 15 min and turned to pale yellow and was then allowed to reach the rt and kept for 1 h. The crude product was filtered off through a celite® pad and rinsed with THF. After evaporation of the THF, the crude was purified via chromatography on silica gel column (petroleum ether/DCM, 9/1, Rf = 0.41 and 0.36). 2 g (68 % yield) of *cis/trans* of **5** were obtained as a beige powder.

**¹H NMR** (300 MHz, Chloroform-*d*) δ 7.79 – 7.69 (m, 2H), 7.68 – 7.57 (m, 3H), 7.53 – 7.39 (m, 2H), 7.34 (d, *J* = 8.4 Hz, 4H), 7.14 (dd, *J* = 8.7, 2.7 Hz, 5H), 6.77 (dd, *J* = 12.1, 8.3 Hz, 2H), 6.59 (dd, *J* = 12.1, 5.9 Hz, 2H).

The spectral data for this compound match those reported in the literature[4].

*11,14-dibromohexahelicene (**Br-[6]H-Br**) (6)*

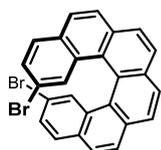

Stilbene **5** (0.25 g, 0.51 mmol, 1 eq) and iodine (8 mg, 0.03 mmol, 0.06 eq) were dissolved in toluene (650 mL) and THF (2 mL). The solution was bubbled with air for 15 min, and then was irradiated under stirring for 16 h with a Hg lamp (150 W). The synthesis was replicated in two batches, for a total amount of 1.2 g of stilbene compound. After evaporation of toluene, the crude was purified by chromatography over silica gel column (petroleum ether/DCM, 9/1, Rf = 0.56). 159 mg (64 % yield) of (*rac*)-**Br-[6]H-Br** were obtained as a light-yellow powder.

The spectral data for this compound match those reported in the literature[5].

The racemic compound was separated into its (*M*) and (*P*) enantiomers by chiral HPLC (*vide infra*).

*(P)-(S, S'-(hexahelicene-11,14-diyl)-diethanethioate) (**7**)*

Synthesized according to the following procedure[5].

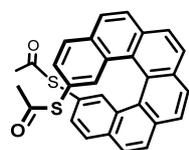

In a microwave flask under argon were dissolved (*P*)-**Br-[6]H-Br** (80 mg, 0.165 mmol, 1 eq), potassium thioacetate (56 mg, 0.49 mmol, 3 eq), xantphos (12 mg, 21 µmol, 0.13 eq) and Pd$_2$.dba$_3$ (9 mg, 10 µmol, 6 mol%) in dioxane (3.5 mL) followed by the addition of freshly distilled Hunig's base (0.1 mL, 0.6 mmol, 2 eq). After degassing with argon, the red solution was irradiated under microwave at 160 °C for 1 h. The organic layer was extracted with ethyl acetate, washed with water, dried over Na$_2$SO$_4$ and concentrated under vacuum. The crude oil was purified by chromatography over silica gel column (petroleum ether/EtOAc, 8/2, Rf = 0.6 and 0.3). 78 mg of (*P*)-**7** (quantitative yield) were obtained as a yellow solid.

**7**

**¹H NMR** (300 MHz, Chloroform-*d*) δ 8.07 – 7.93 (m, 10H), 7.88 (d, *J* = 8.3 Hz, 2H), 7.61 (d, *J* = 1.6 Hz, 2H), 2.15 (s, 6H).

**¹³C NMR** (76 MHz, Chloroform-*d*) δ 194.40, 133.87, 133.53, 132.33, 131.91, 131.44, 129.88, 128.45, 127.94, 127.72, 127.66, 127.63, 127.26, 124.76, 30.01.

**MS (EI) m/z** = 476.0895

The preparation of the (*M*) enantiomer is identical, starting from (*M*)-**Br-[6]H-Br**.

*(P)-hexahelicene-2,15-dithiol (**8**)*

Synthesized according to the following procedure[6].

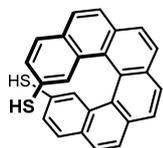

In a 100 mL flask under argon was dissolved (*P*)-**7** (40 mg, 0.084 mmol, 1 eq) in degassed MTBE (22 mL). Then a solution of CsOH.H$_2$O (225.50 mg, 1.34 mmol, 16 eq) in degassed methanol (0.66 mL) was added dropwise and the mixture was stirred for 5 min at rt (the mixture turns yellow). Finally, was slowly added a solution of HCl (3.36 mL, 3.36 mmol, 1M, 40 eq) in H$_2$O (the mixture turns colorless). After extraction with MTBE, drying over Na$_2$SO$_4$ and concentration under vacuum, 32 mg (quantitative yield) of (*P*)-**8** were obtained as a yellow solid.

**¹H NMR** (300 MHz, Chloroform-*d*) δ 8.00 (d, *J* = 7.6 Hz, 4H), 7.89 (d, *J* = 0.6 Hz, 4H), 7.74 (d, *J* = 8.3 Hz, 2H), 7.54 (d, *J* = 1.8 Hz, 2H), 7.22 (dd, *J* = 8.3, 1.9 Hz, 2H), 2.87 (s, 2H).

**¹³C NMR** (76 MHz, Chloroform-*d*) δ 135.93, 133.23, 132.05, 130.22, 130.19, 128.29, 127.58, 127.42, 127.39, 126.66, 126.25, 124.08.

**MS (EI) m/z** = 392.0691

The preparation of the (*M*) enantiomer is identical, starting from (*M*)-**7**.

Chiral HPLC

Analytical chiral HPLC separation for compound **6**

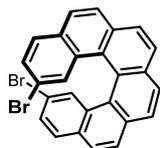

The sample is dissolved in dichloromethane, injected on the chiral column, and detected with an UV detector at 254 nm. The flow-rate is 1 mL/min.

| Column | Mobile Phase | t1 | k1 | t2 | k2 | α | Rs |
|---|---|---|---|---|---|---|---|
| (*S,S*)-Whelk-O1 | Heptane/dichloromethane 80/20 | 6.59 | 1.24 | 7.87 | 1.67 | 1.35 | 5.22 |

**Table S1:** Analytical chiral HPLC separation conditions and characteristics for compound **6**.

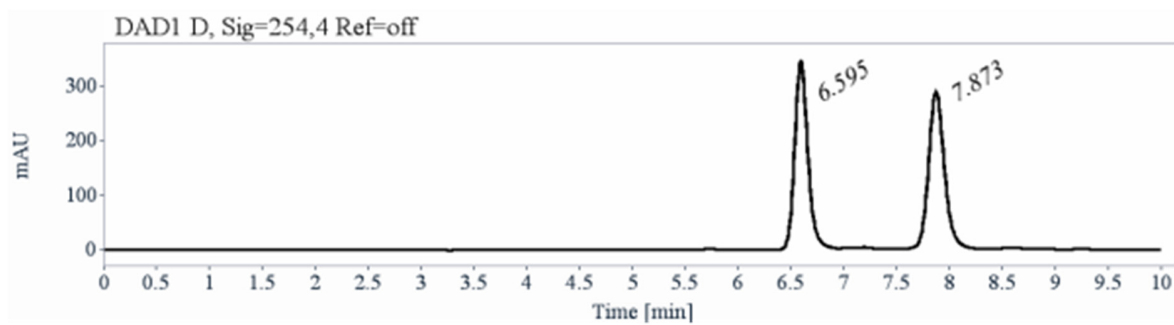

**Fig. S1:** Analytical chiral HPLC for compound **6**.

| RT [min] | Area | Area% | Capacity Factor | Enantioselectivity | Resolution (USP) |
|---|---|---|---|---|---|
| 6.59 | 2964 | 49.91 | 1.24 | | |
| 7.87 | 2974 | 50.09 | 1.67 | 1.35 | 5.22 |
| Sum | 5938 | 100.00 | | | |

**Table S2:** Analytical chiral HPLC separation results for compound **6**.

Semi-preparative separation for compound **6**:

• Sample preparation: About 300 mg of compound **6** are dissolved in 60 mL of dichloromethane.

• Chromatographic conditions: (*S,S*)-Whelk-O1 (250 x 10 mm), hexane / dichloromethane (80/20) as mobile phase, flow-rate = 5 mL/min, UV detection at 350 nm.

• Injections (stacked): 335 times 180 mL, every 4.2 minutes.

- First fraction: 110 mg of the first eluted with ee > 99.5%

- Second fraction: 120 mg of the second eluted with ee >98 %

- Chromatograms of the collected fractions:

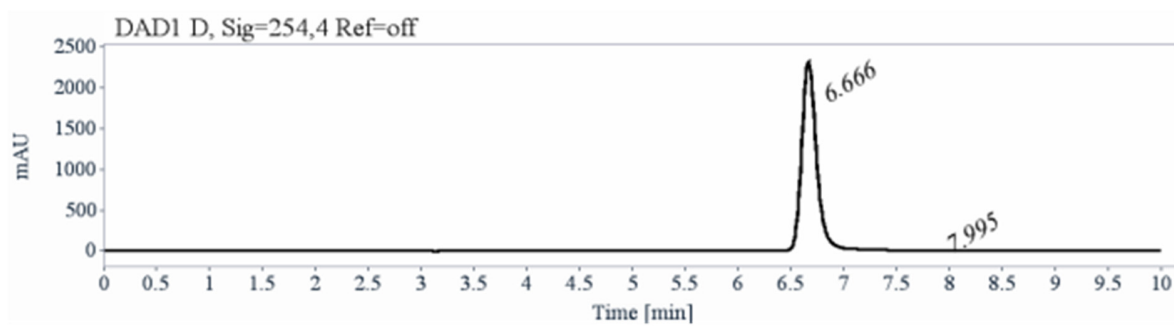

**Fig. S2:** Chiral semi-preparative HPLC separation for compound **6** first eluted.

| RT [min] | Area | Area% |
|---|---|---|
| 6.67 | 21990 | 99.98 |
| 7.99 | 4 | 0.02 |
| Sum | 21995 | 100.00 |

**Table S3:** Semi-preparative chiral HPLC separation results for compound **6** first eluted enantiomer (*P*).

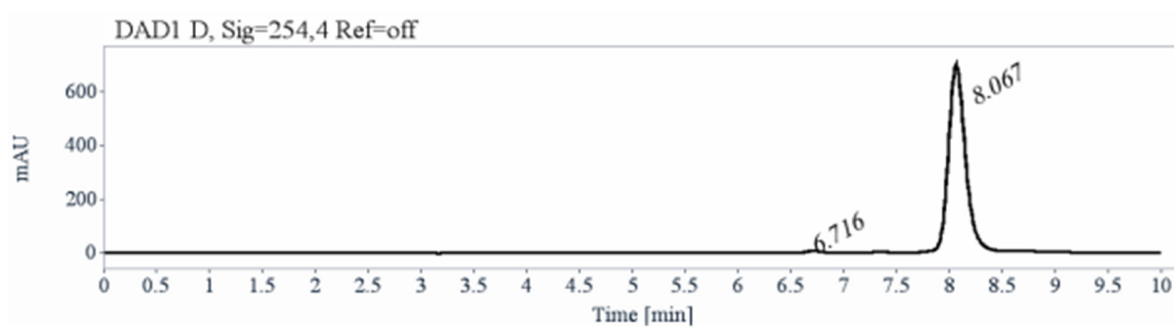

**Fig. S3:** Chiral semi-preparative HPLC separation for compound **6** second eluted.

| RT [min] | Area | Area% |
|---|---|---|
| 6.72 | 67 | 0.84 |
| 8.07 | 7848 | 99.16 |
| Sum | 7915 | 100.00 |

**Table S4:** Semi-preparative chiral HPLC separation results for compound **6** second eluted enantiomer (*M*).

Optical rotations

Optical rotations were measured on a Jasco P-2000 polarimeter with a sodium lamp (589 nm), a halogen lamp (578, 546 and 436 nm), in a 10 cm cell, thermostated at 25°C with a Peltier controlled cell holder.

| $\lambda$ (nm) | **6** first eluted on (*S*,*S*)-Whelk-O1 $[\alpha]_\lambda^{25}$ (CH$_2$Cl$_2$, c =0.037) | **6** second eluted on (*S*,*S*)-Whelk-O1 $[\alpha]_\lambda^{25}$ (CH$_2$Cl$_2$, c =0.038) |
|---|---|---|
| 589 | + 3500 | - 3500 |
| 578 | + 3700 | - 3700 |
| 546 | + 4600 | - 4600 |
| 436 | + 13900 | - 13900 |

**Table S5:** Optical rotations for (*P*)-**6** (first eluted) and (*M*)-**6** (second eluted).

Electronic Circular Dichroism

ECD and UV spectra were measured on a JASCO J-815 spectrometer equipped with a JASCO Peltier cell holder PTC-423 to maintain the temperature at 25.0 ± 0.2°C. A CD quartz cell of 1 mm of optical pathlength was used. The CD spectrometer was purged with nitrogen before recording each spectrum, which was baseline subtracted.

The baseline was always measured for the same solvent and in the same cell as the samples.

The spectra are presented without smoothing and further data processing.

(*P*)-**6**, first eluted on (S,S)-Whelk-O1: green solid line, concentration = 0.158 mmol.L-1 in dichloromethane.

(*M*)-**6**, second eluted on (S,S)-Whelk-O1: red dotted line, concentration = 0.153 mmol.L-1 in dichloromethane.

Acquisition parameters: 0.1 nm as intervals, scanning speed 50 nm/min, band width 1 nm, and 1 accumulation per sample.

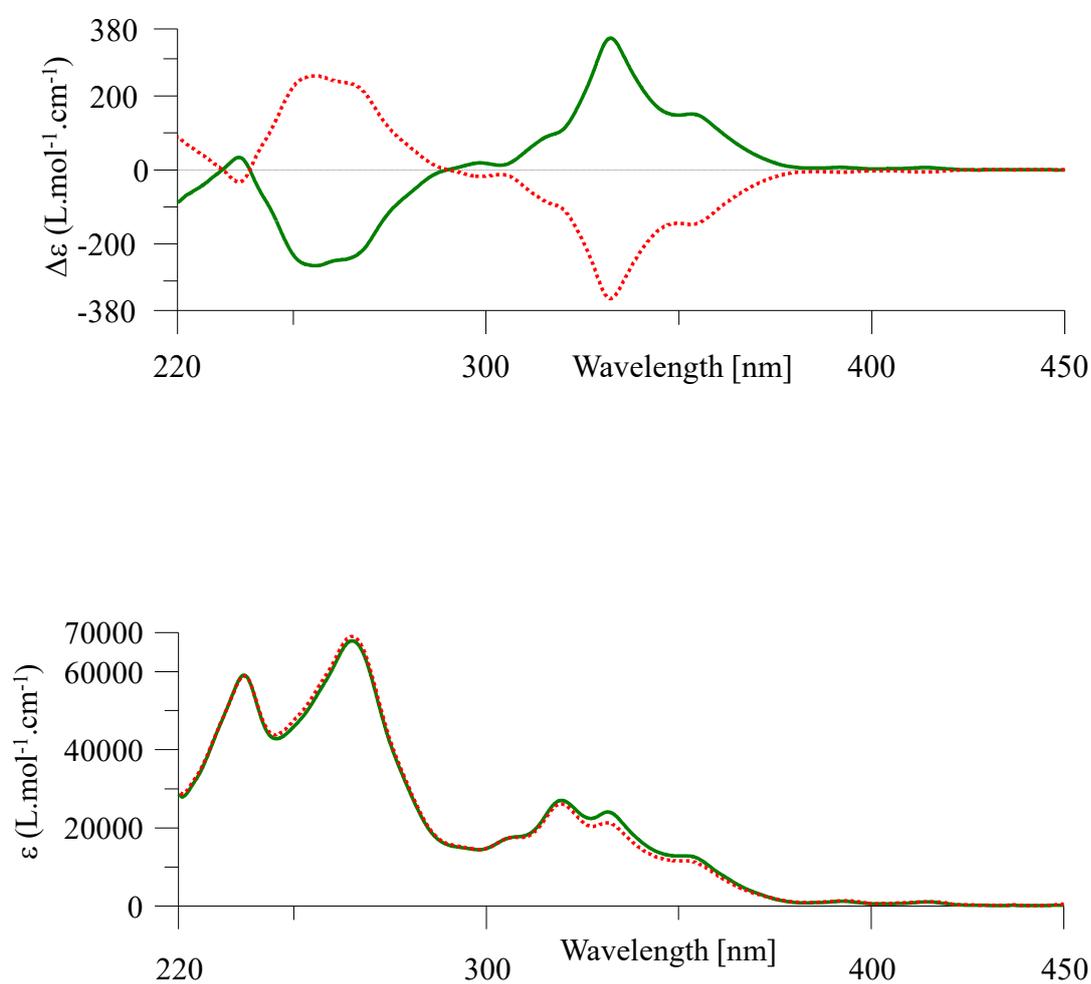

**Fig. S4:** CD (top) and UV-Vis (bottom) spectra of **6** first eluted (green line) and **6** second eluted (red dotted line).

**Section 2: Electron transport measurements of the studied atomic and molecular junctions**

Figures S5,a-c present conductance histograms based on repeated measurements (10,000) of conductance during junction elongation at 200 mV applied voltage for Cu-Cu, Ag-Ag, Au-Au metallic junctions, each with peaks identifies the most-probable conductance during junction stretching. For example, the dominant peak at ~$1G_0$ represents the most probable conductance of a single atom contact[7-9]. As shown in ref. 10, when one metal electrode is made of a softer metal than the other, the softer metal wets the harder

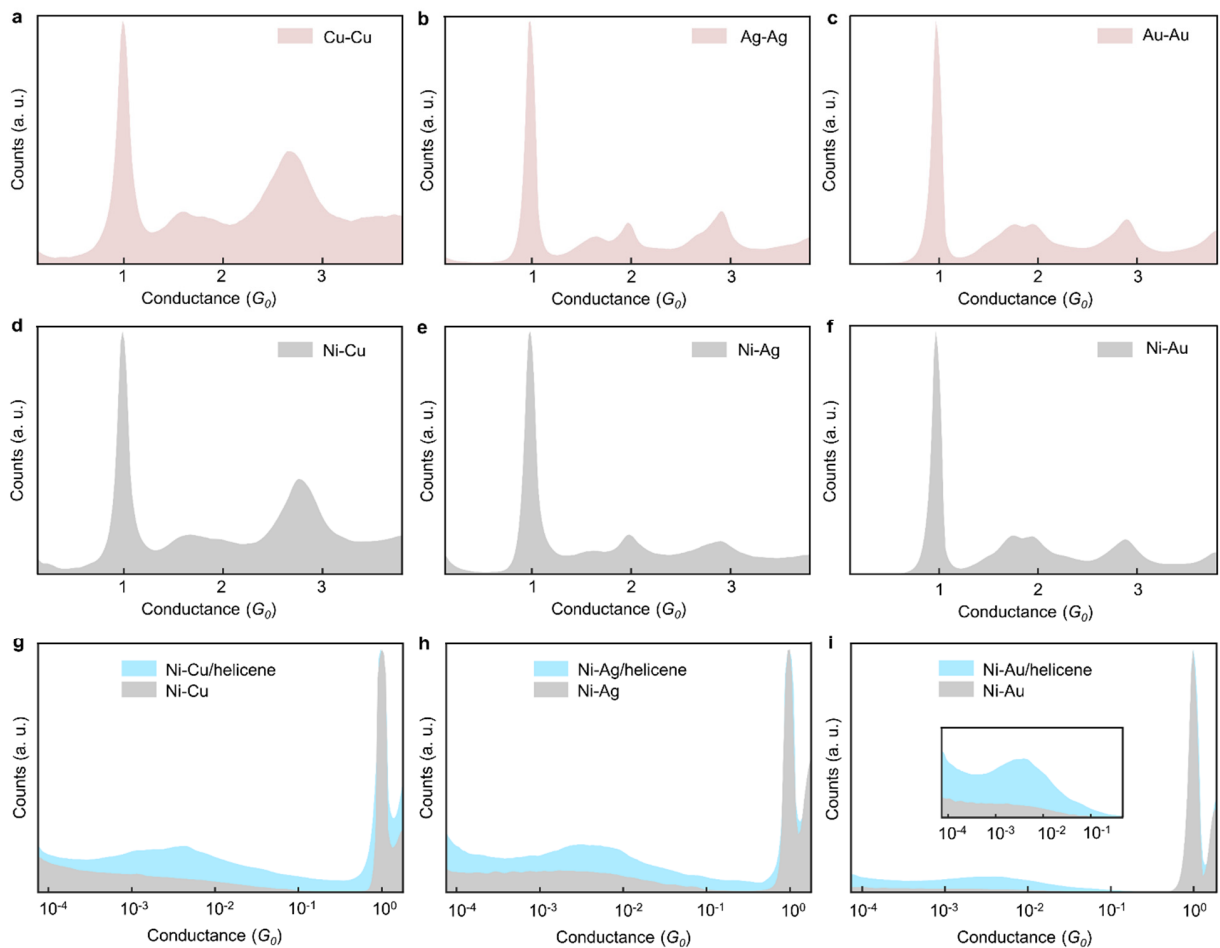

**Fig. S5: Conductance histograms for the considered atomic and molecular junctions. a-i**, Conductance histograms for: Cu-Cu monometallic atomic junctions (a), Ag-Ag monometallic atomic junctions (b), Au-Au monometallic atomic junctions (c), Ni-Cu bimetallic atomic junctions (d), Ni-Ag bimetallic atomic junctions (e), Ni-Au bimetallic atomic junctions (f), Ni-Cu bimetallic atomic junctions in gray and Ni(Cu)-helicene-Cu molecular junctions in light blue (g), Ni-Ag bimetallic atomic junctions in gray and Ni(Ag)-helicene-Ag molecular junctions in light blue (h), and Ni-Au bimetallic atomic junctions in gray and Ni(Au)-helicene-Au molecular junctions in light blue (i). Inset of i: zoom-in image of the blue peak region.

electrode tip, leading to metallic junctions with a constriction made of the soft metal. Similarly, in the studied cases here, the histograms for the Ni-Cu, Ni-Ag, and Ni-Au junctions presented in Fig. S5,d-e are essentially identical to the histograms taken for the Cu-Cu, Ag-Ag, and Au-Au junctions (Figs. S5, a-c), indicating that although the two macroscale electrodes are made of different metals, the atomic-scale constriction within the junctions that dominates their conductance is made of Cu, Ag, and Au respectively. Following the fabrication of metallic junctions with a repeated conductance of the softer metal, the target molecules were introduced. Figure S5,g-f show the conductance histograms of the formed molecular junctions after the introduction of the 2,2'-bis(thiol)-[6]helicene. In all three cases, the most probable conductance is centered between $10^{-3}$ $G_0$ to $10^{-2}$ $G_0$, allowing us to compare ensembles of I-V curves with a similar conductance for all the three cases.

Figure S6 presents the same data as in Fig. 1 in the main text, but in a polar presentation. Figure S7 shows individual I-V curves that were used to construct the average I-V curve and related histograms. To facilitate the observation of individual I-V curves in Fig. S7, we present every 10<sup>th</sup> curve out of the ensembles used in constructing Fig. 1 and Fig. S6. Importantly, across all the measured individual I-V curves, we never observed an asymmetry opposite to that demonstrated by the average curves, for a given chirality and magnetic field.

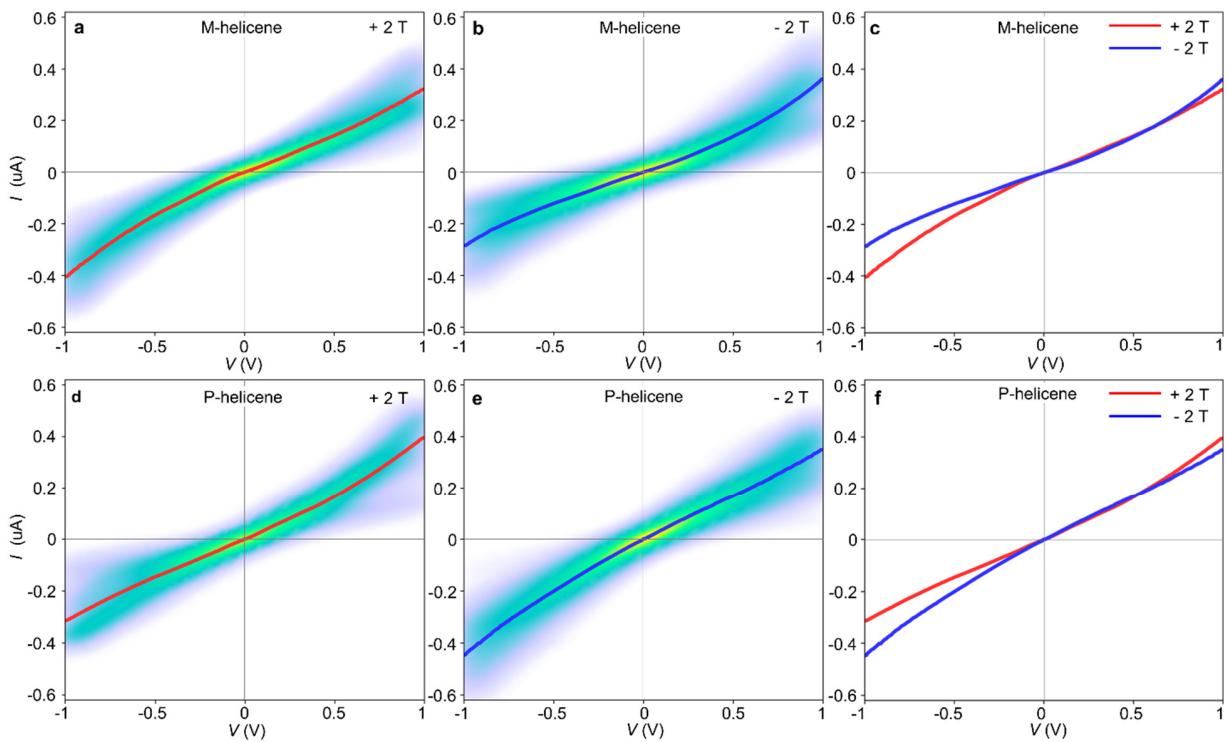

**Fig. S6: Polar presentation for current-voltage analysis of helicene molecular junctions under magnetic fields.**
**a**, Histogram and an average of current as a function of voltage (I-V curves) for Ni(Au)/*M*-helicene/Au junctions under +2T magnetic field, parallel to the junction. **b,** The same under -2T magnetic field antiparallel to the junction. **c,** Average current as a function of voltage for Ni(Au)/*M*-helicene/Au junctions under parallel and antiparallel +2 T and -2 T magnetic fields. **d-f**, The same as (a-c) but for Ni(Au)/*P*-helicene/Au junctions. The number of examined molecular junctions in each case varies between 251 to 377.

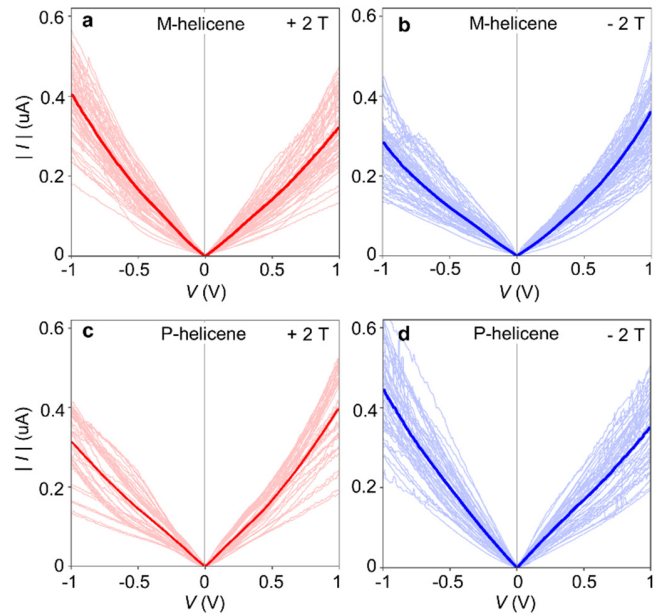

**Fig. S7: Individual I-V curves for the two enantiomers and magnetic field orientations. a**, Current as a function of voltage (I-V) curves for Ni(Au)/*M*-helicene/Au junctions under +2 T magnetic field, parallel to the junction. **b,** The same under -2 T magnetic field antiparallel to the junction. **c-d**, The same as (a-b) but for Ni(Au)/P-helicene/Au junctions. The thick curves represent the average I-V curves for the entire ensembles used in Fig. 1 and Fig. S6. Here, we present every 10$^{th}$ curve out of the mentioned ensembles to allow the observation of individual curves.

## Section 3: Complementary data and Fit procedure in Fig. 2.

Complementary data for Fig. 2 with helicene P-enantiomers

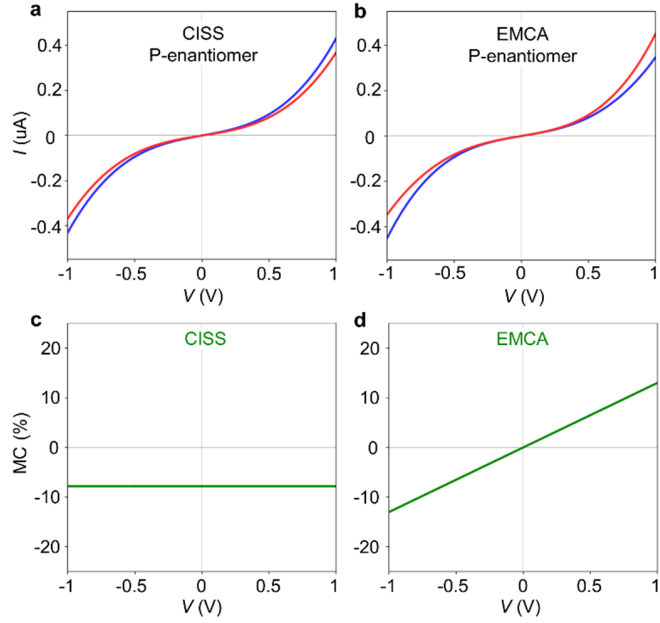

**Fig. S8: Expected I-V curves and MC considering the CISS and EMCA effects for *P*-enantiomers. a,b,** Simulated I-V curve for the CISS effect (a), and EMCA effect (b). **c,d,** Simulated MC for the CISS effect (c), and EMCA effect (d).

Fit procedure in Fig. 2.

As observed in former studies, the I-V curves related to the CISS and EMCA effects are manifested, as described in Fig. 2,a,b and Fig. S8,a,b for the two enantiomers[11-20]. To describe the MC behavior observed in our experiments, we adapted the following approach. For a given molecule's chirality, the total current $I'$ can be expressed as a combination of the intact current $I$, as well as the CISS and EMCA terms, as follows:

$$I' = I + \alpha_i^c I + \beta^c (I \cdot B) V$$

where $I = AV + DV^3$ for a tunneling junction[21]. The constants $A$ and $D$ are determined by the junction's conductance. The constant $\alpha_i^c$ (here, $i = \uparrow, \downarrow$ - electron's spin orientation ; $c$ - chirality) represents the strength of the CISS effect for a given chiral conductor, where its sign depends on the electron's spin orientation, $i$ (parallel or antiparallel to the conductor), and the conductor's chirality. The term $\alpha_i^c I$ effectively reproduces the previously observed I-V behavior related to the CISS effect. The last term, $\beta^c (I \cdot B) V$, describes the EMCA effect. It relies on both the current $I$ and the magnetic field B[22]. The constant $\beta^c$ signifies the strength of the EMCA effect for a given chiral conductor, and its sign depends on the molecule's chirality.

For the *M*-enantiomer, Figs. 2,a,b illustrates the I-V curves attributed to the CISS and EMCA effects, respectively. In a parallel magnetic field orientation (B>0) and a positive voltage, the EMCA term induces a negative correction to the current (see red curve in Fig. 2b). For the same magnetic field and voltage conditions, the CISS term elicits a positive correction (see red curve in Fig. 2a). The latter is attributed to the prevalence of spins oriented antiparallel to the junction's axis when the Nickel electrode's magnetization is parallel, as a result of a dominant minority spin population at the Fermi energy. In contrast, for a negative voltage, both terms contribute positively, leading to an enhanced current magnitude. The situation is inverted for an antiparallel magnetic field orientation (B<0). To correctly describe the contributions of the CISS and EMCA effects for the *P*-enantiomer, the signs of $\alpha_i^c$ and $\beta^c$ should be inverted.

Figures 2, C and D depict the MC contributions attributed to the CISS and EMCA effects, respectively, in view of Figs. 2,a,b and the MC definition given in the main text. The fit procedure is described as follows. The experimentally obtained data in Fig. 2g (black curve) is fitted to a linear function, ignoring fine details that may be ascribed to the local junction's electronic structure. The intercept with the MC axis defines the MC in Fig. 2c and the slope defines the curve's slope in Fig. 2d. Once the MC curves in Figs. 2,c,d are obtained and in view of the MC expression, the values of A, D, $\alpha_i^c$, and $\beta^c$B (for B=+2 T or -2 T) were found, and the I-V curves presented in Figs. 2,a,b were constructed.

Remarkably, we only changed the signs of $\alpha_i^c$ and $\beta^c$, which were extracted by the fit to the experimental MC data for the *M*-helicene junctions (Fig. 2g) to obtain the linear curve (green) in Fig. 2h that fits very well the experimental MC data (black), found for *P*-helicene junctions in independent experiments.

The used values for the fitting parameters are A=1· $10^{-7}$ µAmp/V, D=3· $10^{-7}$ µAmp/V³. For the M-enantiomer and a parallel magnetic field (+2 T), we used $\alpha_i^c$= 0.078, and $\beta^c$= -0.0625 $T^{-1}V^{-1}$. For the M-enantiomer and an antiparallel magnetic field (-2 T), we used the same $\alpha_i^c$ magnitudes but with an opposite sign, while $\beta^c$ remains the same (and $\beta^c$B changes sign due to the change in B sign). To generate Fig. S8 for the case of junctions based on the *P*-enantiomer, we simply took the opposite signs for $\alpha_i^c$ and $\beta^c$, with identical A and D parameters.

## Section 4: The relation between asymmetry and conductance difference.

We define asymmetry as:

$Asymmetry = 100 \cdot [|I(+V)| - |I(-V)|]/[|I(+V)| + |I(-V)|]$

Dividing the numerator and denominator by $V$, where: $V = |(+V)| = |(-V)|$ is the voltage magnitude for the same positive and negative voltage, $(+V) = -(-V)$,

we get: $Asymmetry = 100 \cdot [|I(+V)|/V - |I(-V)|/V]/[|I(+V)|/V + |I(-V)|/V]$

For: $G(+V) = |I(+V)|/V$, and $G(-V) = |I(-V)|/V$,

we have: $Asymmetry = 100 \cdot [|G(+V)| - |G(-V)|]/[|G(+V)| + |G(-V)|]$

Since the conductance difference is defines as: $\Delta G(V) = |G(+V)| - |G(-V)|$,

we get: $Asymmetry = 100 \cdot [\Delta G(V)/[|G(+V)| + |G(-V)|]]$

Namely, the asymmetry is proportional to the conductance difference, for a given voltage magnitude.

## Section 5: Magnetoconductance shift extracted by linear fitting

In the analysis presented in Fig. S9, we ignore the fine structure of the MC curves and use a linear fit to evaluate the MC shift at zero voltage in order to have a quantity that is not solely determined by the specific MC obtained at zero voltage.

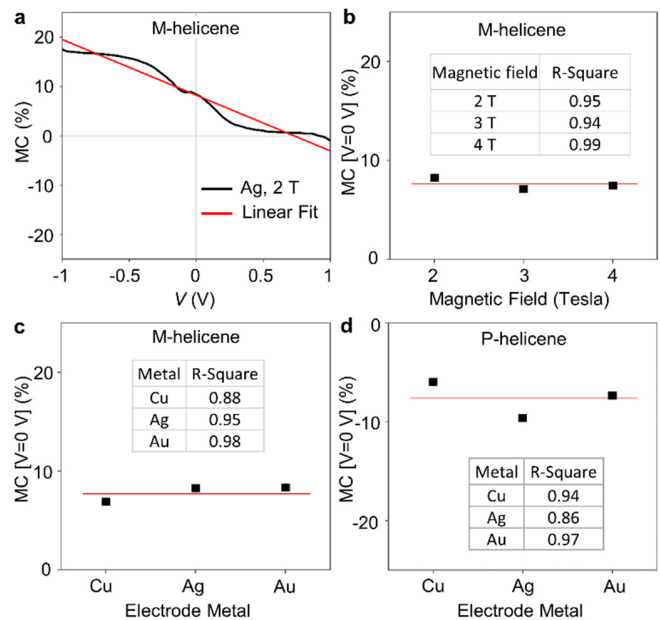

**Fig. S9: Magnetoconductance shift analysis based on linear fits. a,** Example of a linear fit (red) to a MC curve (black). **b**, MC shift based on the linear fit value at zero voltage as a function of magnetic field magnitudes for junctions based on *M*-helicene. **c**, MC shift based on the linear fit value at zero voltage as a function of the counter electrode metal type for junctions based on *M*-helicene. **d,** The same as (c) for junctions based on P-helicene. Error bars are smaller than the symbols. The R-Square of each fit is presented in the inset tables.

## Section 6: Additional complementary figures

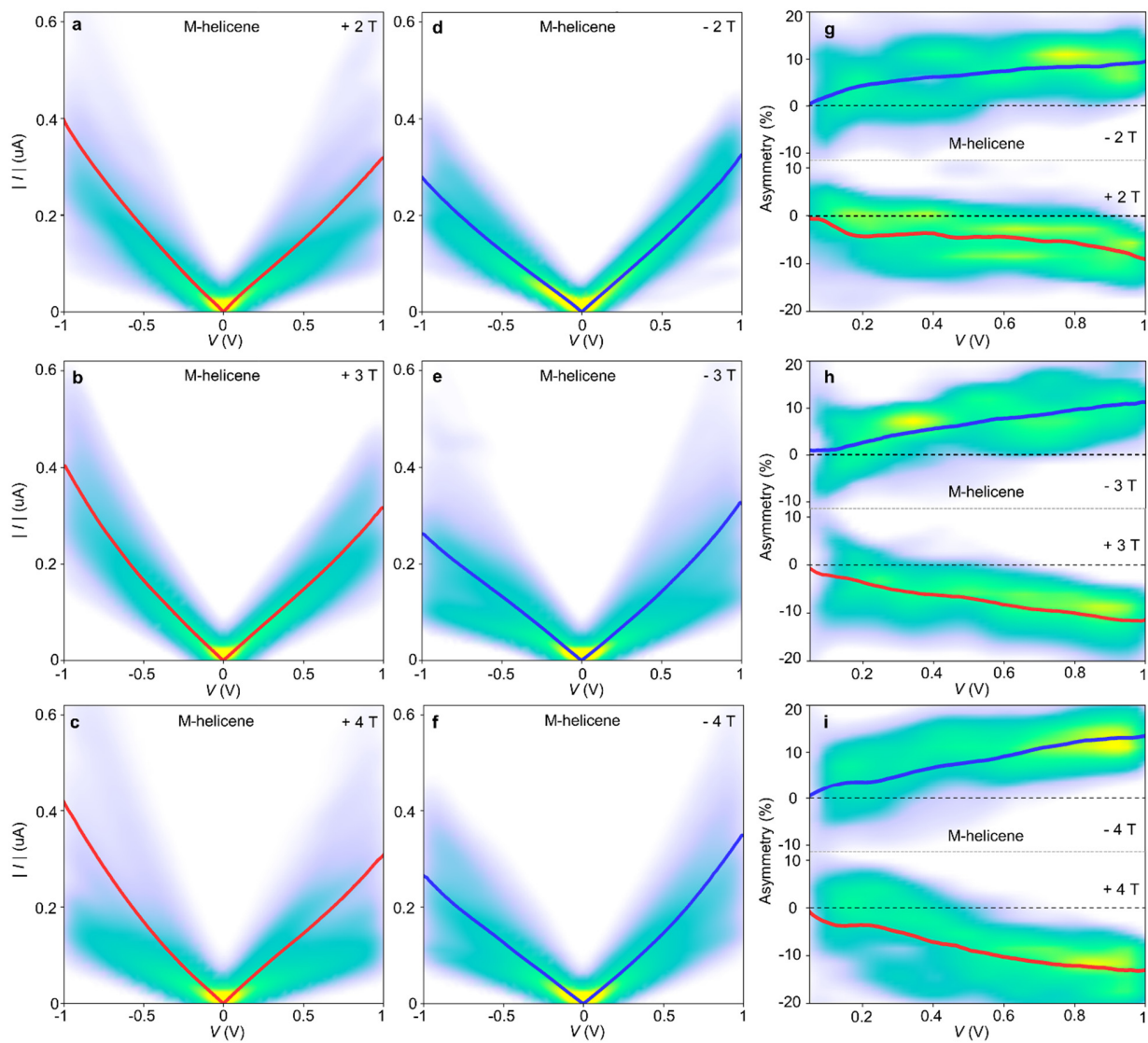

**Fig. S10: Current as a function of voltage and asymmetry histograms at different magnetic fields. a-f,** Histogram and an average of current in absolute values as a function of voltage for Ni(Ag)/*M*-helicene/Ag junctions at different applied magnetic fields. **g-l,** Histogram and an average Asymmetry as a function of applied voltage magnitude at different applied magnetic fields. The number of examined molecular junctions in each case varies between 372 to 634.

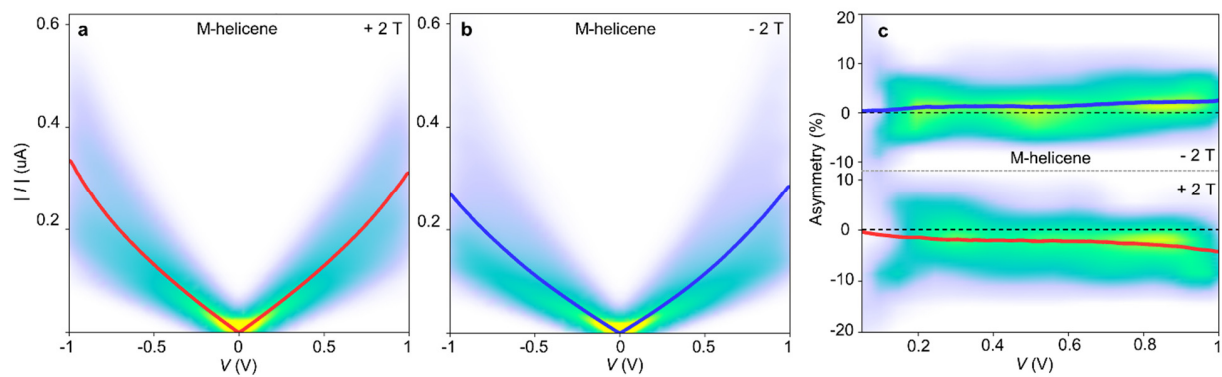

**Fig. S11: Current as a function of voltage and asymmetry histograms for Cu based junctions. a,b**, Histogram and an average of current in absolute values as a function of voltage for Ni(Cu)/*M*-helicene/Cu junctions at parallel and antiparallel magnetic field orientations. **c**, Histogram and an average Asymmetry as a function of applied voltage magnitude. Measurements were done at an applied magnetic field of +2 T or -2 T. The standard error of the current is smaller than the curve width. The number of examined molecular junctions is 416 for +2 T and 443 for -2 T.

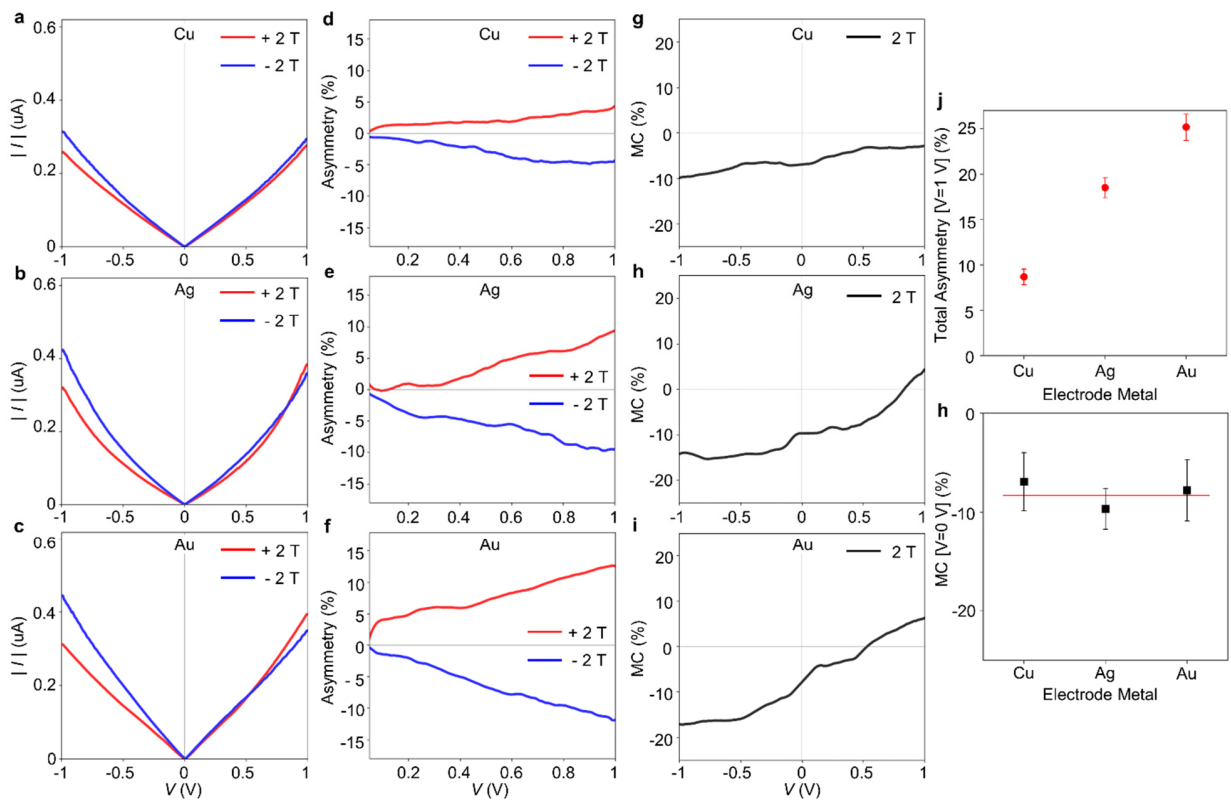

**Fig. S12: The response of asymmetry and MC to different metal electrodes for *P*-helicene based junctions. a-c,** Average current (in absolute values) as a function of applied voltage for Ni(X)/*P*-helicene/X junctions, where X is Cu (a), Ag (b) and Au (c). The standard error of the current is smaller than the curve width. **d-f,** Average asymmetry as a function of applied voltage magnitude for the same junctions as in (a) to (c), respectively. **g-i,** Average MC as a function of applied voltage for the same junctions as in (a) to (c), respectively. **j,** Total asymmetry at 1 V (sum of positive and negative asymmetry magnitudes) for junctions based on different metals. **k,** MC shift (MC at zero voltage; black squares) for junctions based on different metals. The red curve represents the average value. The number of examined molecular junctions in each case varies between 248 and 448. The error bars for asymmetry and MC indicate the experimental uncertainty. See Fig. 1,g,h,k and Fig. S13, for |I|-V and asymmetry histograms for the Au, Cu, and Ag based junctions.

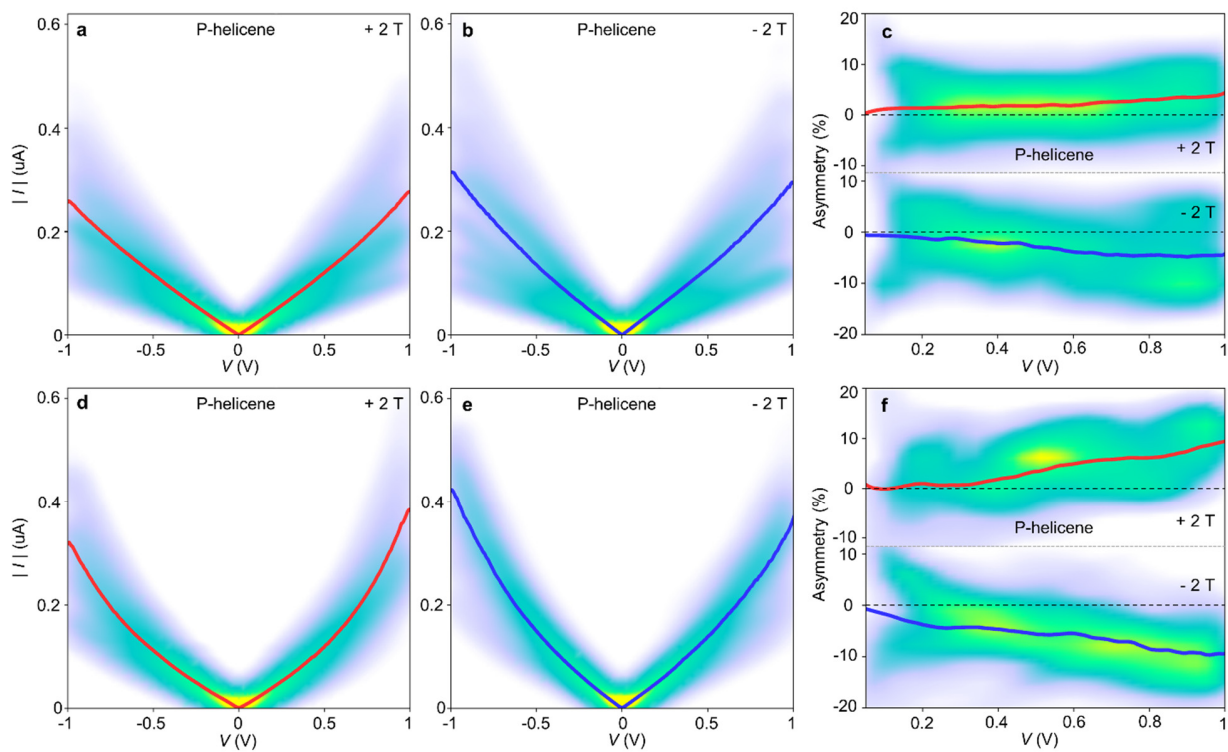

**Fig. S13: Current versus voltage and asymmetry histograms for junctions based on *P*-helicene with different electrode metals. a,b,** Histogram and an average of current in absolute values as a function of voltage for Ni(Cu)/*P*-helicene/Cu junctions at parallel and antiparallel magnetic field orientations. The standard error of the current is smaller than the curve width. **c,** Histogram and an average Asymmetry as a function of applied voltage magnitude at parallel and antiparallel magnetic field orientations for Ni(Cu)/*P*-helicene/Cu junctions. **d,e,** The same as (a,b) but for Ni(Ag)/*P*-helicene/Ag junctions. The standard error of the current is smaller than the curve width. **f,** The same as (c) but for Ni(Ag)/*P*-helicene/Ag junctions. Measurements were done at an applied magnetic field of +2T or -2 T. The number of examined molecular junctions in each case varies between 248 to 448.